\documentclass[12pt,preprint]{aastex}

\shorttitle{Ly$\alpha$ emitters at $0.2 < z < 0.35$}
\shortauthors{Deharveng et al.}

\begin{document}


\title{Lyman $\alpha$ emitting galaxies  at $0.2 < z < 0.35$ \\
    from GALEX spectroscopy}


\author{
Jean-Michel Deharveng\altaffilmark{1},
Todd Small\altaffilmark{2},
Tom A. Barlow\altaffilmark{2}, 
C\'eline P\'eroux\altaffilmark{1},
Bruno Milliard\altaffilmark{1},
Peter G. Friedman\altaffilmark{2},
D. Christopher Martin\altaffilmark{2},
Patrick Morrissey\altaffilmark{2},
David Schiminovich\altaffilmark{3},
Karl Forster\altaffilmark{2},
Mark Seibert\altaffilmark{2},
Ted K. Wyder\altaffilmark{2},
Luciana Bianchi\altaffilmark{4},
Jose Donas\altaffilmark{1},
Timothy M. Heckman\altaffilmark{5},
Young-Wook Lee\altaffilmark{6},
Barry F. Madore\altaffilmark{7},
Susan G. Neff\altaffilmark{8},
R. Michael Rich\altaffilmark{9},
Alex S. Szalay\altaffilmark{5},
Barry Y. Welsh\altaffilmark{10}, 
Sukyoung K. Yi \altaffilmark{6},
}

\altaffiltext{1}{Laboratoire d'Astrophysique de Marseille, BP 8, Traverse
du Siphon, 13376 Marseille Cedex 12, France}

\altaffiltext{2}{California Institute of Technology, MC 405-47, 1200 East
California Boulevard, Pasadena, CA 91125}

\altaffiltext{3} {Department of Astronomy, Columbia University, New York, NY
  10027}

\altaffiltext{4}{Center for Astrophysical Sciences, The Johns Hopkins
University, 3400 N. Charles St., Baltimore, MD 21218}

\altaffiltext{5}{Department of Physics and Astronomy, The Johns Hopkins
University, Homewood Campus, Baltimore, MD 21218}

\altaffiltext{6}{Center for Space Astrophysics, Yonsei University, Seoul
120-749, Korea}

\altaffiltext{7}{Observatories of the Carnegie Institution of Washington,
813 Santa Barbara St., Pasadena, CA 91101}

\altaffiltext{8}{Laboratory for Astronomy and Solar Physics, NASA Goddard
Space Flight Center, Greenbelt, MD 20771}

\altaffiltext{9}{Department of Physics and Astronomy, University of
California, Los Angeles, CA 90095}

\altaffiltext{10}{Space Sciences Laboratory, University of California at
Berkeley, 601 Campbell Hall, Berkeley, CA 94720}









\begin{abstract}
    The {\it GALEX (Galaxy Evolution Explorer)} spectroscopic survey mode,
    with a resolution of  $\sim$ 8 \AA \ in the  FUV (1350 - 1750 \AA) and $\sim$ 20 \AA \ in
    the NUV (1950 - 2750 \AA) is used
    for a systematic search of  Ly$\alpha$ emitting galaxies at low redshift.
  This aims at filling a gap between high-redshift surveys and 
      a small set of objects studied in detail in the nearby universe.
 A blind search of 
 7018 spectra extracted in 5 deep exposures (5.65 sq.deg) has resulted in
   96  Ly$\alpha$ emitting galaxy candidates in the FUV domain,
    after accounting for broad-line AGNs.
   The  Ly$\alpha$ EWs (equivalent
    width) are consistent with stellar population model predictions and show
    no trends as a function of UV color or UV luminosity, except a possible 
    decrease in the most luminous that may be due to small-number statistics.
    Their distribution in EW
    is similar to that at z  $\sim$ 3 but their fraction among star-forming
    galaxies is smaller. Avoiding uncertain candidates, a sub-sample of
    66 objects
    in the range $0.2 < z < 0.35$  has been used to build 
    a  Ly$\alpha$ LF (luminosity function). The incompleteness due to objects  
    with significant  Ly$\alpha$ emission but a UV continuum too low for
    spectral extraction has been evaluated. A comparison with  H$\alpha$ LF 
    in the same redshift domain is consistent with an average
    Ly$\alpha$/H$\alpha$  of $\sim$ 1 in about 15 \% of the star-forming
    galaxies.
    A comparison with high-redshift  Ly$\alpha$ LFs implies an 
    increase of the  Ly$\alpha$ luminosity density by a factor of about 16 
    from  z  $\sim$ 0.3 to  z  $\sim$ 3. By comparison with the factor 5
     increase 
    of the UV luminosity density in the same redshift range,
   this suggests an increase  of the average  Ly$\alpha$ escape fraction 
   with redshift. 
\end{abstract}





\section{Introduction}

  The Lyman $\alpha$ emission line has attracted large attention as a spectral signature
  for identifying galaxies and securing redshifts at large distances. This emission line,
  however, is  fraught with difficulty.
  Because Ly$\alpha$ photons are resonantly scattered by neutral hydrogen, 
  they may suffer more dust attenuation than adjacent UV continuum photons; 
  their escape is also affected by the relative geometries of  neutral 
  and ionized interstellar gas, and, last but not least, by the velocity structure of neutral gas.
  The first two factors have been extensively discussed in the light of 
   IUE observations of nearby star-forming galaxies \citep{gia96} (and references therein) and 
   with  model calculations  \citep{cha93,neu91,che94} (and
  references therein).
     The crucial role of 
      the velocity structure of neutral gas has been shown 
       by the  HST  spectra of nearby star-forming galaxies
  \citep{kun98,mas03}. Similar evidence was offered by the 
    spectra of Lyman Break Galaxies (LBG) \citep{pet98,pet00}.
       Ly$\alpha$ photons mainly escape when they are scattered off
  neutral gas that is offset in velocity from the bulk of the ionized regions. The  complexity of the escape
   of  Ly$\alpha$ emission is also well illustrated by 
the broad distribution of  Ly$\alpha$ strengths and profile-types observed 
in the LBG spectroscopic  sample of  \citet{sha03}. 
The Ly$\alpha$ transmission mechanisms, especially  the resulting emergent line profiles, 
have since been investigated in increasingly realistic models (e.g.)  \citep{ahn01,ahn02,
 ahn03,han06,ver06}. An extensive 
  review of all the aspects of the observations of the  Ly$\alpha$ emission line in galaxies
  has been recently given by \citet{scha07}.

        The complex nature of Ly$\alpha$  escape was  advocated  for the disappointing 
        results  of earlier searches of distant Ly$\alpha$ emitters
        (e.g.) \citep{djo92}. Nonetheless,
       the Ly$\alpha$ emission remains the only 
mean for identifying galaxies when the continuum becomes too faint to be detected,
and, 
 following \citet{cow98} and  \citet{hu98}, 
increasingly deeper and larger surveys
have come into a widespread use for detecting galaxies at high
        redshifts.
 Beyond the redshift of $\sim 6$, 
 the increasingly neutral IGM is not  a complete 
obstacle to the visibility of Ly$\alpha$ emission (e.g.)  \citep{hai02}
and  the density evolution of 
Ly$\alpha$ emitters may even help to trace the history 
 of the cosmic re-ionization (e.g.)  \citep{mal04,ste05,kas06}.

Although nearby galaxies have played a key role 
for understanding the factors affecting  Ly$\alpha$ escape, their observations,
using space-borne UV spectrographs in pointing mode, 
were directed to specific and known objects.  As a consequence,
 there  is  not yet a  systematic survey for redshifts smaller than those reachable from the ground.
 We use here
    the {\it GALEX} spectroscopic survey mode for the first  systematic search 
    of  Ly$\alpha$ emitting galaxies at low redshift.  
    Goals are to understand (i) whether the   Ly$\alpha$ escape is related to  specific properties 
    of galaxies, 
  (ii) whether the  Ly$\alpha$ emission evolves 
 from current epoch to high z as 
   the cosmic star formation rate traced by Balmer lines or the 
  UV continuum of galaxies. 
If so,   the average relationship between the massive 
stellar content of the galaxies and the  Ly$\alpha$ emission  would be constant
over time;
 the  Ly$\alpha$ emission might be used as a  tracer of star formation,
with an empirical  calibration  encapsulating the average effects of resonant scattering.
If not, there would be an evidence for cosmic evolution of the
physical processes, especially galactic winds, expected to play a central 
role  in the  Ly$\alpha$ escape from galaxies.


\section{Data analysis}

  \subsection{{\it GALEX} ultraviolet spectroscopy and selection of line emitters}


      The {\it GALEX} instrument and mission are described by \citet{mar05} and \citet{mor05}. 
The spectroscopy mode utilizes a CaF$_2$ grism that can be moved into the
convergent beam of the telescope 
to form simultaneous spectra of all sources in the field in both FUV and NUV
bands. The usable wavelength range for relatively faint sources is
approximately 1350 to 1750 \AA \ for FUV (2nd order) and 1950 to 2750 \AA \ for
NUV (1st order). The  
spectral resolution for a point source (assuming a 5 arcsecond
PSF) is  $\sim$ 8 \AA \ for FUV and $\sim$ 20 \AA \ for NUV.
Details on the observations (mutiple grism orientations in order to avoid
the spectra overlap) and  the various steps of data reduction are given by
\citet{mor07}.
The spectral extraction is performed for all 
 point sources in the direct image observations exceeding a S/N limit per   
resolution element in  the co-added results of 2 in FUV and 3 in NUV.
Each resulting spectrum is a one-line image of 488 pixels
with 3.5 \AA \ per pixel and  starting at 1300 \AA.  

       Eleven fields observed in spectroscopy mode are available from the {\it GALEX} GR2. 
In order to keep a relative homogeneity in the detection depth  and 
 related selection effects we have concentrated on 
   five fields with an exposure time 
larger than 70,000 s and covering an area of 5.65 square deg.  The characteristics of these five fields are summarized in
 Table 1, with the total number of spectra extracted by the reduction pipeline
and the number of objects identified in the same fields with the direct 
imaging mode.  
Each spectrum has been visually inspected and  potential 
 Ly$\alpha$ emission features have been measured  (central wavelength,
 line flux, equivalent width and full width at half maximum, FWHM)
 with the IRAF splot package (gaussian fitting). The signal to noise ratio
 does not permit reasonable identifications of  Ly$\alpha$ absorption. 

The objects with a FWHM larger than about 15 \AA\  in the FUV domain
 and about 27 \AA\  in the NUV are classified as broad-line AGNs.
 These  limits are based on the minimum value of 1200 km s$^{-1}$, observed
 in the distribution of the FWHM of H$\alpha$ emission line 
 of galaxies in the SDSS  \citep{hao05a} and found to make a separation between  the
broad-line AGNs and other emission-line objects. These limits take into account 
  the average redshift and the spectral resolution in the respective GALEX UV bands. 
The presence/absence of  emission lines such as OVI, CIII, CIV, 
when the redshift is appropriate and the brightness large enough for detection in the GALEX 
bands, is also used to confirm the classification, especially when the FWHM
 values are close to the limit values.

  In the NUV domain, only broad-line AGNs are found according to our criterions. This is consistent 
  with the NUV (AB) limiting magnitude of 22 of the vast majority of extracted
  spectra.
 At the redshift of 0.65, the lowest redshift at which a  Ly$\alpha$ emitter can be detected in the NUV band, 
  this flux limit would imply an (AB) absolute magnitude 
brighter than $-21$,  a value extreme and rare for galaxies, as shown by the evolution 
of the galaxy 1500 \AA \  luminosity function \citep{arn05}.
 This situation is aggravated by the dilution of narrow spectral features into
 the  $\sim$ 20 \AA \  NUV spectral resolution which plays against the
  detection of Ly$\alpha$ emitting galaxies.

It is found impossible to identify narrow-line AGNs  among our potential Ly$\alpha$ emitting galaxies
  because either their associated CIV line is redshifted into the noisy
  wavelength domain
 between the FUV and NUV, or the object is too faint for a detection of 
 the  CIII or CIV lines. A contamination by narrow-line AGNs is probably present
 and it will be seen in the next subsection whether additional spectral information can help.

\begin{figure}
\plotone{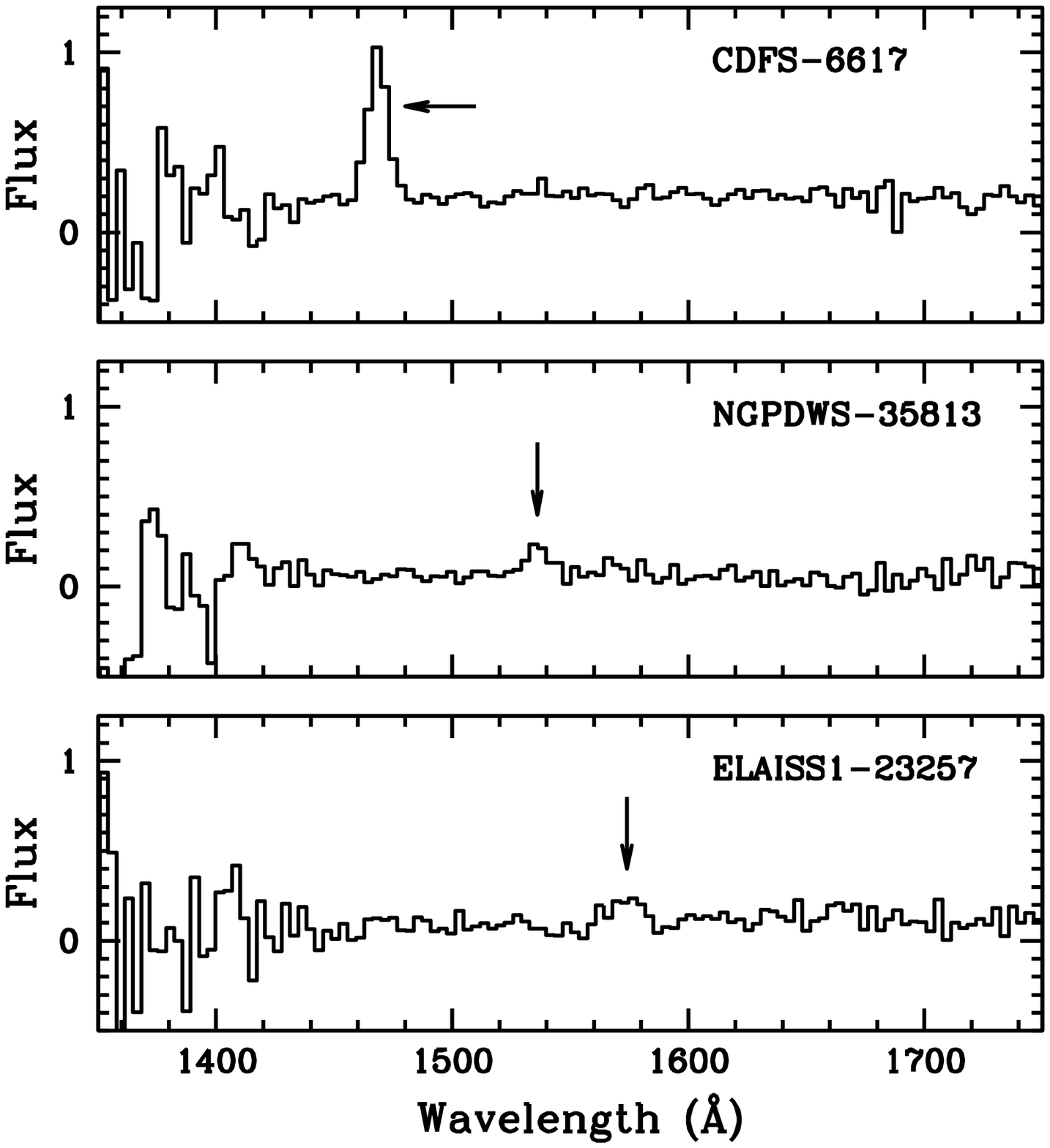}
\caption{ Examples of {\it GALEX} spectra (in $10^{-4}$ photons cm$^{-2}$  s$^{-1}$
A$^{-1}$ flux unit) and features identified as Ly$\alpha$ emission with qualities
good (Q = 1), fair (Q = 2), uncertain (Q = 3) 
 from top to bottom.
\label{fig1}}
\end{figure}

  As our blind search was open to all extracted spectra and not limited to galaxies with
   the appropriate redshift, we have an increased risk of spurious detections.
  We have therefore classified our candidates into  three categories
  (good, fair and uncertain) in order to try to control if the properties of our objects depend on the 
  quality  of their identification, even though the sample sizes are changing.
 Fig. \ref{fig1} gives a few examples of these spectra with the proposed identification.
    The identification of potential Ly$\alpha$ features is also hampered
  at both ends  of the spectral range by 
fluctuations that increase  because of the
decrease in efficiency. In between, 
the useful spectral range is not constant  from
object to object. For the problems that require a
control of the volume surveyed, we have defined
 the largest wavelength domain that we have been able to  search in relatively uniform conditions for
 all the objects. This wavelength range  1459 -- 1642 \AA\ 
 translates  in a redshift window 
0.2 -- 0.35 for the  Ly$\alpha$ emission line. 




 The number of Ly$\alpha$  emission line candidates
  resulting from  our  blind search, as well as those retained  for the discussion of luminosity functions
  (quality 1 and 2 only; $0.2 < z < 0.35$) are listed in Table 1
for each field investigated. The total numbers are 96 and 66 respectively in these two categories.
Table 2 summarizes the main characteristics of our candidates 
  with the line flux of the emission features, and, 
  assuming an identification to  Ly$\alpha$ emission, the redshift and the derived  Ly$\alpha$ luminosity.
      The line fluxes and  Ly$\alpha$ luminosities include a correction factor resulting from a 
    re-calibration of the spectral response, that was verified to make no systematic 
    differences on average between the direct image photometry and the fluxes derived from 
    an integration of the spectra. The one-sigma precision on the line flux
  measurement is of the order of $4 \times 10^{-16}$ erg cm$^{-2}$  s$^{-1}$. In
  combination with an evaluation of the uncertainty in the
  determination of the UV continuum
  adjacent to the emission line, this number gives a precision on the
  equivalent width (EW)  from about 25 \%
  at EW $\sim$ 20 \AA\ to 15 \% at EW $\sim$ 100 \AA.

\subsection{Cross-verification  with spectroscopic or photometric redshift information} 
     
      As a control of our blind search, we have compared our results with 
      existing information about each candidate, especially  any redshift 
 from spectroscopy or photometry in the optical.  We are mostly concerned by 
spurious features and not confusion  with another emission line since
      Ly$\alpha$  
emission is known as  unique  in the FUV  spectra of galaxies. 
       The possibility of identifying narrow-line AGNs would also be of interest since we have shown  
       it was difficult in practice with the UV spectra alone.
      
   Each of our 96 candidates was searched in the NED (NASA/IPAC Extragalactic
 Database) and the recent Data Release 3 of DEEP2  \citep{dav07}; 
 the 14 redshifts found are displayed in Table 2. Only one of these redshifts, 
 based on COMBO-17 photometric measurements  \citep{wol04} and possibly
 affected by misidentification in a dense group,
    is in significant disagreement with the evaluation based
   on the Ly$\alpha$  emission. 
   All 14 objects are classified as galaxies. Although this cross-verification
   is presently limited to a small subset of candidates, 
   it is seen as an encouraging validation of our approach.           
   The small number of  redshifts available from the visible was somewhat expected: on one hand,   the 
      SDSS spectroscopic survey (available in 3 of our 5 fields) has a small fraction of galaxies
       at $z  >  0.2$,  on the other hand the {\it GALEX} spectroscopy is not deep enough 
  for overlapping well the specific and deep redshift surveys with large telescopes.

\section{Properties of the  Ly$\alpha$ emitting galaxies} 

\subsection{Comparison with  the UV-selected galaxy population} 

     Our  Ly$\alpha$ emitting galaxy candidates 
 are compared with the other UV sources of {\it GALEX}, especially those 
    identified as galaxies, in order to see whether  the presence of the Ly$\alpha$ line   
    is related to  any galaxy  property.  Such a comparison can also
    illustrate
    the importance of  selection effects.
    In the plot of the FUV flux vs. UV color,  the 96  Ly$\alpha$ emitting candidates lie 
    at  the faint magnitude and blue color boundaries of the domain occupied by 
    the 736 objects classified as galaxies but without  Ly$\alpha$ features.
    The faint magnitude boundary  has a natural explanation with 
our candidates being faint since  in principle at $z > 0.2$.
    The blue color boundary  has two possible explanations.
 On one hand, easier  Ly$\alpha$ escape may come with less dust and bluer
    color; 
 on the other hand, bluer color may imply a higher FUV continuum flux and easier 
    detection of emitting features.

\begin{figure}
\plotone{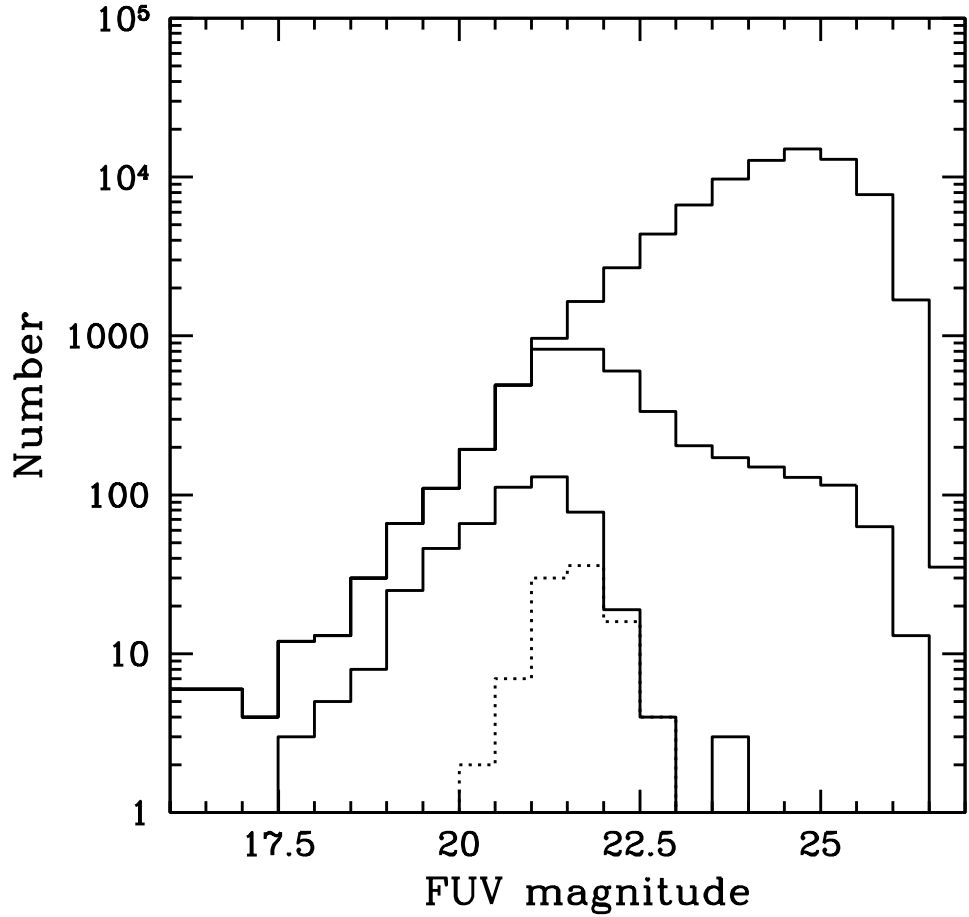}
\caption{Distribution of FUV (AB)magnitudes (all five fields selected). From top 
to bottom (solid line) sources detected in the images, sources with an extracted 
spectrum and  sources classified as galaxies. The dotted line is for the  Ly$\alpha$
emitting galaxy candidates.
\label{fig2}}
\end{figure}

    Another aspect of the comparison  of our  Ly$\alpha$ emitting candidates
    with respect to the general population of {\it GALEX} sources  is shown  in Fig. \ref{fig2}
    with  the distribution of the FUV magnitudes. This distribution is displayed  for four samples, 
     the UV sources detected in the images,  the extracted spectra, the objects classified as galaxies
     (without any emission feature)
     and  the Ly$\alpha$ emitting  galaxy candidates. The 
    spectra appear to have been systematically extracted down to a magnitude of 21.5, 
    a level at which samples are  complete
    according to the completeness analysis of {\it GALEX} images by \citet{xu05}.
    The distribution of objects classified as galaxies also peak at this
    limit. 
    At the bright end of the distribution, the proportion of 
    galaxies with Ly$\alpha$  emission is naturally low since the objects 
    are selected at $z > 0.2$. At the faint end, the number of objects classified as galaxies and the 
    number of  Ly$\alpha$ emitting galaxy candidates are very similar. This does not mean that all 
   identified  faint galaxies have Ly$\alpha$  in emission since the samples are by construction distinct.  
   This tells that the identification of faint galaxies is relatively easier with than without an emission 
   feature. The distribution of Ly$\alpha$ emitting galaxy candidates  
     peak at $21.5 - 22$ magnitude. They are clearly affected by incompleteness.

\subsection{ Distribution of Ly$\alpha$ EW}      

\begin{figure}
\plotone{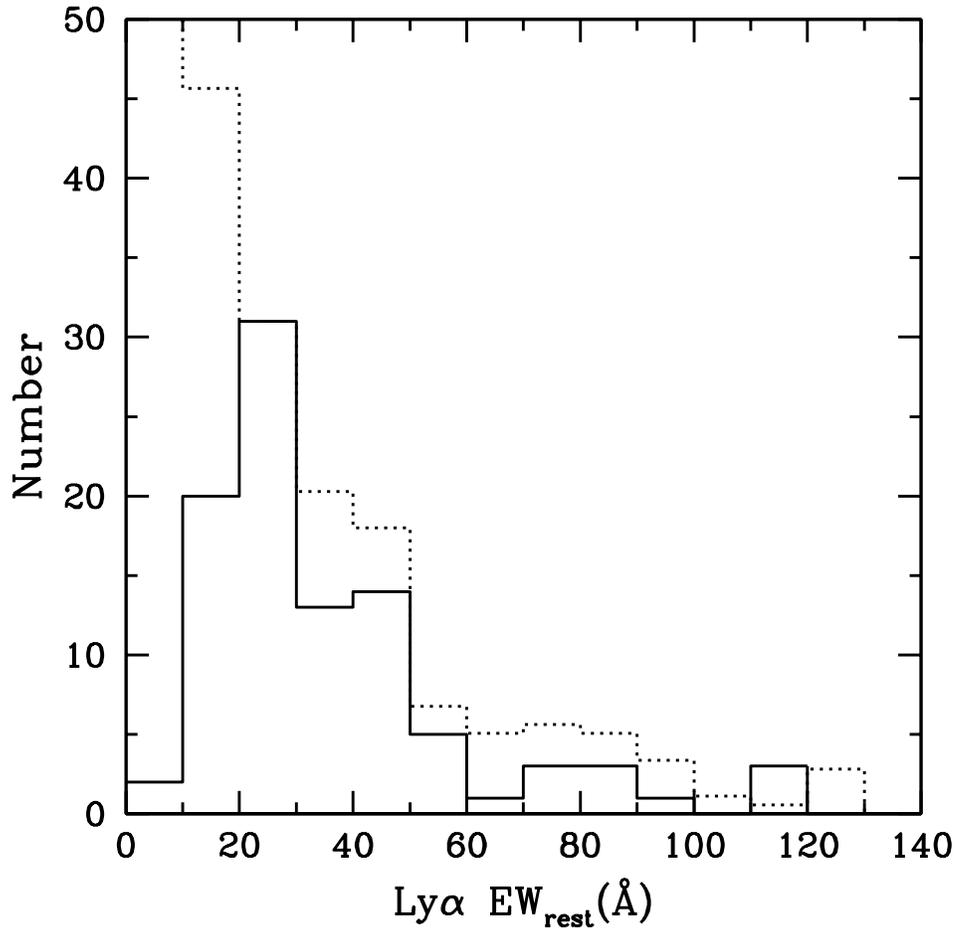}
\caption{ Distribution of Ly$\alpha$ rest-frame equivalent width (EW).   
The  distribution of  \citet{sha03} is shown (dotted line) after a normalisation with our data
in the $20 - 30$  \AA  \ bin.
\label{fig3}}
\end{figure}

    The distribution of  Ly$\alpha$ rest-frame equivalent widths is displayed
    in Fig.  \ref{fig3}. Beyond a completeness limit at about 20 \AA, the
    distribution is comparable  with that of  \citet{sha03} for LBGs 
    at $z \sim 3$. The fraction of  Ly$\alpha$ emitting galaxies (with EW $>
    20$  \AA) relative to the number of star-forming galaxies in the redshift range $0.2 <
    z < 0.35$ cannot be determined directly because redshift
    measurements are available only in limited areas and,
    in these areas, the number of matches 
    with {\it GALEX} spectra is very small. We can rely on evaluations based on the 
    {\it GALEX} far-UV survey and use the luminosity functions derived by  \citet{arn05}
    in the redshift range $0.2 < z < 0.4$ to calculate the total number of
    galaxies up to a given magnitude  in a volume comparable to our observed
    volume. Up to magnitude (AB) of 21 and 21.5, the number of 
    Ly$\alpha$ emitting galaxies with EW $>  20$ \AA\ are respectively 9 (out of 58) and
    36 (out of 243), corresponding to a fraction of 15 \% of the total number of 
    (star-forming) galaxies.
    This fraction is lower than the fraction of 25 \% reported by \citet{sha03} for LBGs 
    at $z \sim 3$.   This difference may reflect differences between the methods of evaluation.
    The spectroscopic sample of  \citet{sha03} does not have a UV flux limit as we had to 
    use for the determination of the size of the parent population since 
     our candidates are searched among objects of unknown redshifts.
     On the other hand, the difference is consistent  with the trend of 
     lower incidence of  Ly$\alpha$ emission at low redshift found by  
     \citet{red07};  they, however, report a fast decline of the fraction of 
      Ly$\alpha$ emitting galaxies (EW $>  20$ \AA) with a value 
      of  8 \% in  the redshift bin $1.9 < z < 2.17$.

    The Ly$\alpha$ rest-frame EWs  displayed in Fig.  \ref{fig3} are consistent
    with the large spread of values predicted by 
     the stellar population models of
    \citet{cha93} and any  amount 
    of  Ly$\alpha$ quenching in the resonant scattering  process.
     The EW values  larger than model predictions that raise problems at high
    redshift  \citep{shi06,fin07,sta07} are not found.

\subsection{ Ly$\alpha$ dependences: {\it GALEX} data}                 

     We also examine how the  Ly$\alpha$ strength varies across our sample as a 
    function of different galaxy parameters. The most obvious parameters are those 
    relative to the UV continuum emission as obtained from {\it GALEX} photometry; these data have the
    advantage to be naturally available for all our candidates.
      
\begin{figure}
\plotone{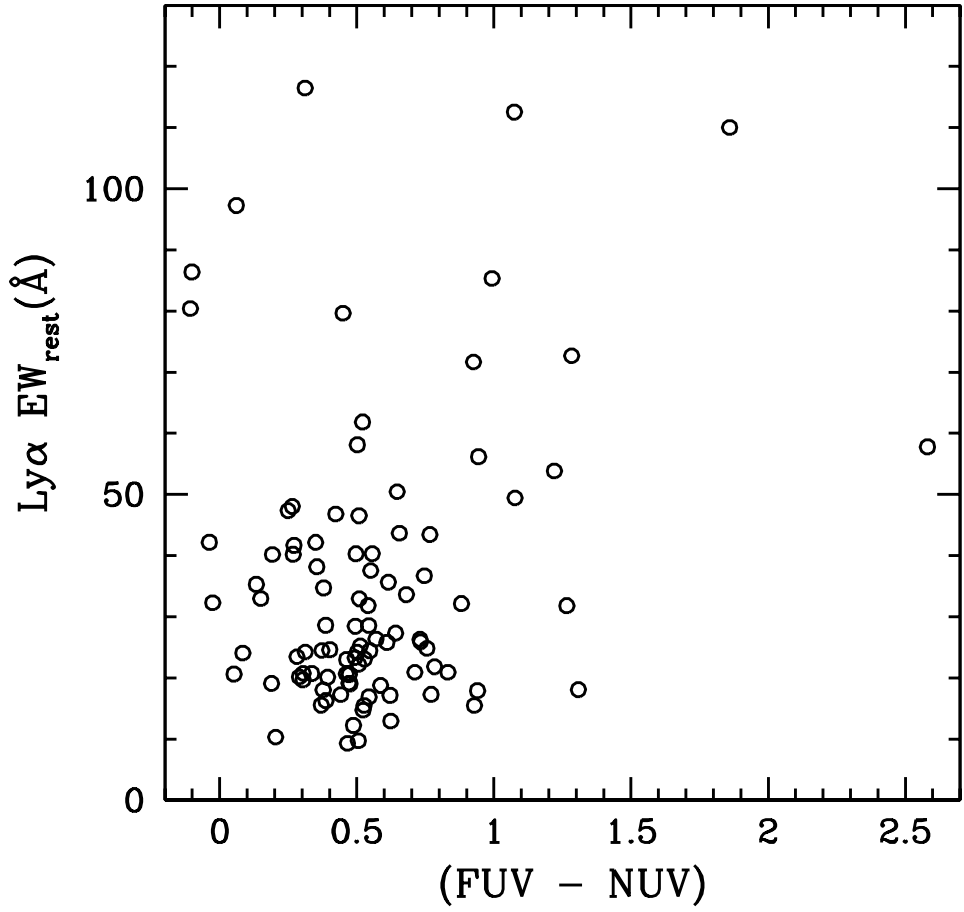}
\caption{ Ly$\alpha$ rest-frame EW as a function of the (FUV$-$NUV) color.
\label{fig4}}
\end{figure}

     The  variation of the Ly$\alpha$ rest-frame EW as a function of the UV color does not show any 
     trend (Fig. \ref{fig4}).  Insofar as the UV color  reflects the continuum extinction,
      this is consistent with a decoupling of the reddening of line and continuum photons
      in  the resonant scattering process.
     This is in contrast with the 
     trend reported by  \citet{sha03} of the EW increasing when the
     UV continuum slope becomes bluer. Their trend, however, encompasses a much wider 
     range of EW  than ours, from
     strong absorption to strong emission, and would be less
     significant if restricted to our actual limited range of EW. 
     
\begin{figure}
\plotone{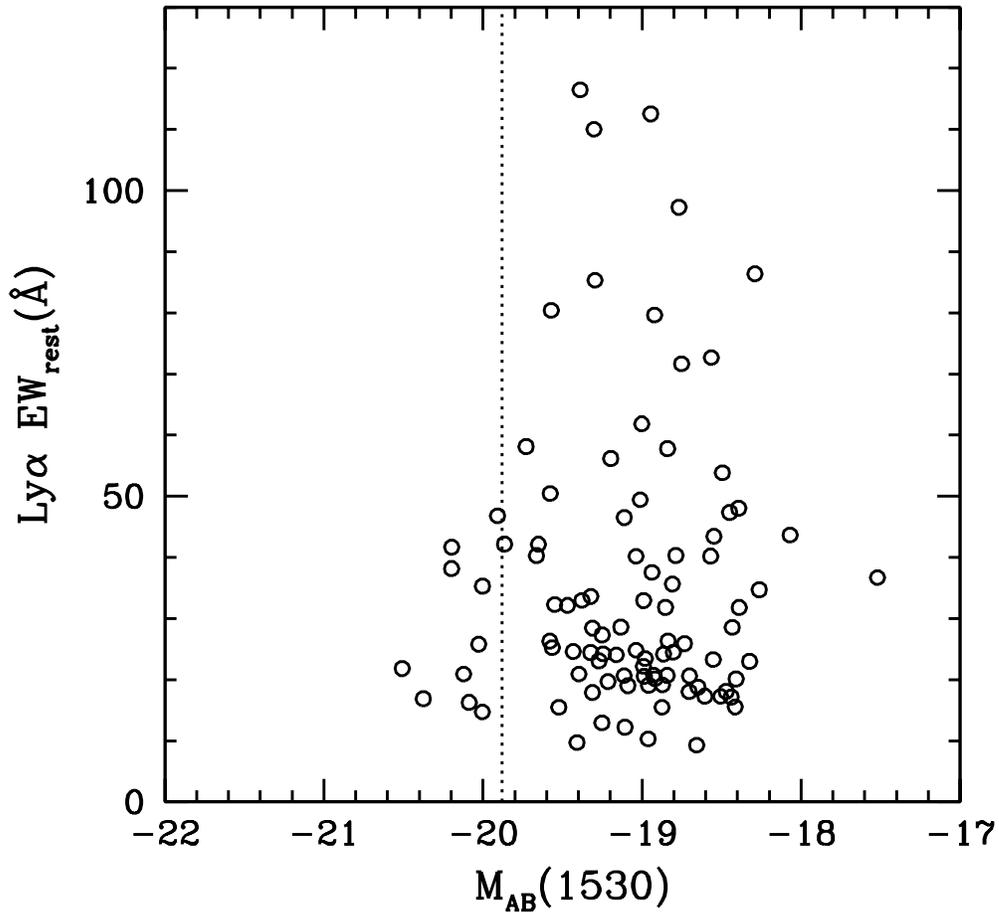}
\caption{ Ly$\alpha$ rest-frame EW as a function of the FUV absolute
  magnitude. The objects brighter than $-$ 19.9 (vertical dotted line)
  are UVLGs according to the definition of  \citet{hec05}. The high-redshift
  galaxies discussed by \citet{and04} are also at the left of this line.
\label{fig5}}
\end{figure}

     The Ly$\alpha$ rest-frame EW does not reveal
      any trend either as a function of the UV luminosity (Fig.  \ref{fig5}). 
      The deficiency of strong Ly$\alpha$  emissions reported in 
      the most luminous high-z Lyman Break Galaxies (LBGs)
      \citep{and04,and06,shi06} 
      is not directly comparable 
      since it takes place 
         at an absolute magnitude $ < -21$, a limit that we do not reach in our
      sample.
      However, the fraction of objects with large EW is small in Fig.  \ref{fig5} and such objects 
       would likely be missing in a smaller sample of Ly$\alpha$ emitters. This
      suggests that, in addition to the interpretations given by  \citet{and04}, 
      the trend seen at high redshift may be due or enhanced by 
      small number statistics.

        Fig.  \ref{fig5}  allows us also to identify
         ultra-violet luminous galaxies (UVLGs). This class of galaxies was  defined 
          by  \citet{hec05} 
         among {\it GALEX} local galaxies 
         to overlap the luminosity range of typical high-z LBGs. 
          With the actual definition of  \citet{hec05}  we have 11 UVLGs  in our sample
          at the left of the dotted vertical line in Fig.  \ref{fig5}. 
          These UVLGs do not have
          EW  as large as those found in less luminous objects. This
         is comparable, albeit at lower 
          luminosity, with the trend reported  for the high-z LBGs. This may also be 
          explained by small number statistics.
                 Using the UV continuum LF obtained
          in the redshift range $0.2 - 0.4$ by  \citet{arn05}, 
        we calculate  a density of  UVLGs of
          about $2 \times 10^{-5}$  Mpc$^{-3}$, predicting 
           about 24 of these objects in the redshift window $0.2 < z < 0.35$ 
         and our 5 fields. With this evaluation, 
           about 33  \% (8/24) of the  UVLGs have
         Ly$\alpha$   in emission (with an EW 
           larger than about 20 \AA) which is a larger fraction than
          found above for the general population. Given the luminosity range
         of the UVLGs and the redshift window, incompleteness cannot explain
         that difference. In contrast the evaluation of the number of luminous 
           galaxies based on the bright end of the UV luminosity function is
         uncertain. On the other hand it is possible that a larger UV
         luminosity may contribute to a larger  Ly$\alpha$ escape. This would
         be consistent, in addition to a possible   evolution of the  Ly$\alpha$
          escape fraction, with the higher incidence 
          of  Ly$\alpha$ emission at higher redshifts 
          reported by  \citet{red07} and  \citet{ouc07}.

\subsection{ Ly$\alpha$ dependences:  corollary information}        

     Galaxy parameters which are not derived from {\it GALEX} observations are 
     available for only a fraction of our relatively distant objects.
     An extreme example is
    the H$\alpha$ line emission which would be interesting in order to get  the amount of ionizing radiation without the complication   
    of resonant scattering  and, therefore, the 
    Ly$\alpha$ escape fraction. In spite of the 12 spectroscopic redshifts found in the
     literature (Table 2), we
     have been unable for various reasons (essentially lack of calibrated
     fluxes) to recover the H$\alpha$ flux in
     more than 2 galaxies. The resulting   Ly$\alpha$/H$\alpha$ ratios are
     found to be 19
     (GROTH-21024) much above the theoretical recombination ratio of 8.7 and 5
     (GROTH-34512) which is larger than any reported value
      in near-by star forming galaxies \citep{gia96}. These findings
     illustrate the difficulties of spectrophotometric comparisons and the
     need for a dedicated optical spectroscopic follow-up.

\begin{figure}
\plotone{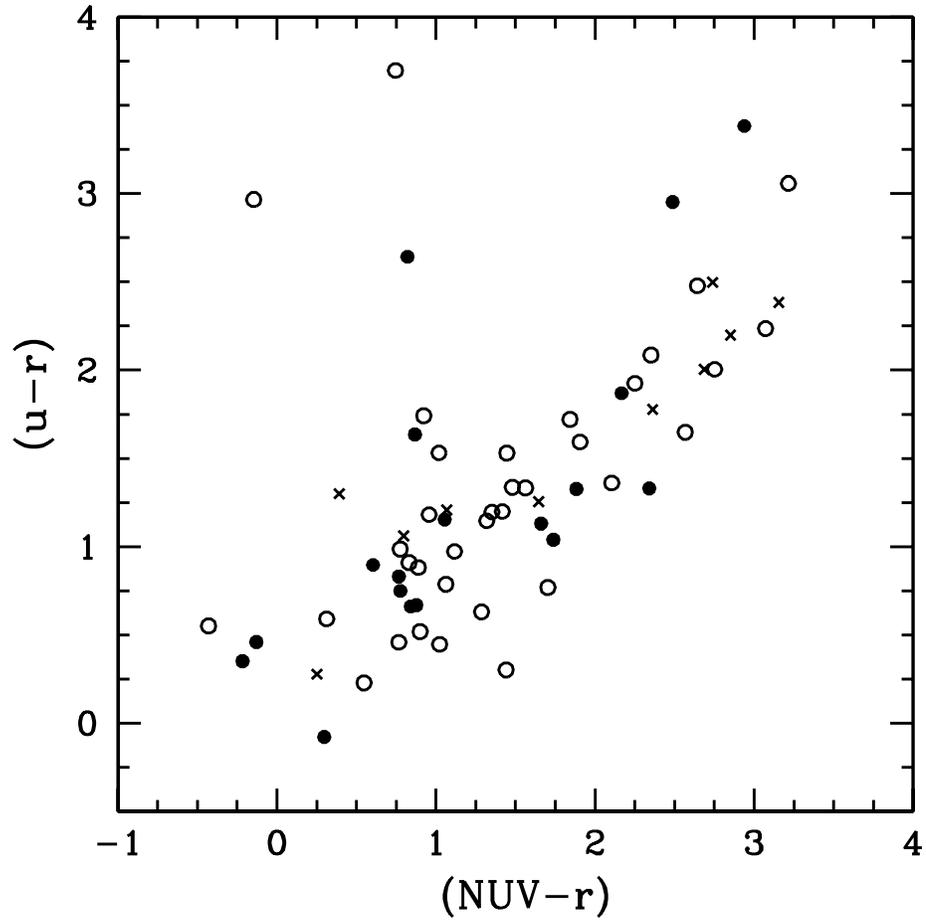}
\caption{ (u-r) vs. (NUV-r) color-color diagram built with 
    the {\it GALEX} NUV flux and
    the SDSS flux measurements available for 64 of our 
      Ly$\alpha$ emitting galaxy candidates. 
      One candidate with a  low quality emission feature and a (u-r) color of
      about 6 out of the range expected from galaxies has been discarded.  Different symbols
      are used according to the  Ly$\alpha$ rest-frame EW: 
      EW $<$ 30 \AA (open circle),  30 $<$ EW $<$ 50 \AA (solid circle),  EW $>$ 50 \AA (crosses).
\label{fig6}}
\end{figure}

   The  SDSS photometry (u, g, r, i, z filters) is available for three of our
   five fields and 64 of our  Ly$\alpha$
    emitting galaxy candidates. 
    We have calculated the (NUV$-$r) color which has been extensively used by  \citet{wyd07} in their
    study of galaxy colors with {\it GALEX} 
   and  is known to make a pronounced demarcation 
    between the blue and red sequences. The range of colors obtained for our  Ly$\alpha$
    emitting galaxy candidates is in good agreement with the values expected from galaxies.
     We find no trend between this color and the
   Ly$\alpha$ rest-frame EW, the Ly$\alpha$ line luminosity and the FUV luminosity.
    The (u$-$r) vs. (NUV$-$r)  color-color diagram  (Fig.  \ref{fig6}) shows a 
      sequence in good agreement
   with the blue part of the sequence (NUV$-$r $ <  $ 4)
   obtained in the same diagram by  \citet{wyd07} on a very large sample of galaxies (their figure 22). In contrast  the density of galaxies along our sequence is relatively constant whereas it increases in the 
   diagram of  \citet{wyd07} 
   with the  (NUV$-$r)  color increasing from 0 to $\sim 3$. This difference results probably from the selection of our objects among  star-forming galaxies with a significant far-UV continuum and at 
   redshift $z > 0.2$.
      We have separated the objects of  Fig.  \ref{fig6}  into three groups according to their
      Ly$\alpha$ EW values. These categories 
     do not appear segregated either along or perpendicular to the 
    sequence which  is  primarily driven, as suggested by  \citet{wyd07},  by
   star formation history.
  Since the  Ly$\alpha$ EW (without any transfer) is predicted rather stable  
    as a function of time in galaxies with constant
    star formation \citep{cha93}, 
   this illustrates the dominant role of individual radiation transfer effects
   in the Ly$\alpha$ escape rather than the 
    stellar population properties (e.g.)  \citep{scha08}

\section{The Ly$\alpha$   luminosity function} 

\subsection{Volume evaluation and space density} 

      For this approach we need  conditions of detection as uniform as
      possible. We have therefore
    restricted ourselves to the redshift range $0.2 - 0.35$ 
     and have retained  only the objects of quality 1 or 2 in
      order to limit  spurious detections. We have used the V/V$_{max}$
      method. 
   For each Ly$\alpha$ emitting galaxy candidate, V$_{max}$ is the volume over
      which a source of the same  Ly$\alpha$ luminosity could lie and still meet
      the blind search criteria.  The inverse volumes of all the galaxies in a
      particular luminosity bin are summed to estimate the luminosity function
      in that bin.

 V$_{max}$ is defined by the field of view
     and the redshift range $0.2 - zl$. The redshift $zl$ is 
       0.35 if the  Ly$\alpha$ luminosity is bright enough that
      the line remains above the line flux limit out to the upper bound of the redshift window.
  The corresponding volume, maximum value of  V$_{max}$, is 
 $2.369 \times 10^5$ Mpc$^{-3}$ taking into account a 0.6 deg field of
      view radius. For fainter sources, $zl$ is
 the redshift ($< 0.35$) at which the  Ly$\alpha$ flux falls
      below the line flux limit. In this determination the
      Ly$\alpha$ flux is decreased as the inverse square of the luminosity
      distance since the  Ly$\alpha$ is not spectrally resolved in the
      galaxies. 
      The determination of $zl$ is somewhat uncertain since it is based 
on the line flux limit which is evaluated empirically in each field and
      results from both the depth of each field and the continuum level of the 
spectra. In practice,  slight adjustements have been made to account 
   for the specific noise in each spectra. Because of the relatively narrow
redshift window, only the sources with a  Ly$\alpha$ luminosity less than 10$^{42}$ 
erg s$^{-1}$ may be affected by these uncertainties in volume evaluation.

       In order to combine the results obtained over the five fields we have to deal 
with the differences of depth. With the upper redshift limit of 0.35, the sources with  
 Ly$\alpha$ luminosity log(L) $>$ 41.8 (erg s$^{-1}$) are essentially above the line
flux limits in all fields and can be reasonably merged. Adopting a bin width
of 0.2 in log(L), we have summed up the inverse volumes in their
respective bins over the 5 fields and divided by 5 to account for the
increased volume. It is possible to get the luminosity function below  the
limit of  log(L) = 41.8 (erg s$^{-1}$) at the expense of cosmic variance
by using only the deepest three fields, CDFS, GROTH and NGPDWS (cf Table 1).
We have repeated the summations of inverse volumes over these three fields. 
The bin below  log(L) = 41.6 (erg s$^{-1}$) is again affected by
incompleteness 
 and the lowest bin with only one source has been discarded as unsignificant.

 The resulting luminosity function 
      per log Ly$\alpha$ luminosity is plotted in  Fig. \ref{fig7}.     
 As a generic consequence of detection thresholds in images,
  the lowest bins of the luminosity function are affected by incompleteness. 
 We have not tried to remedy this type of incompleteness and will only refrain
 to use the lowest luminosity bins in
 further discussion.

\subsection{Correction for incompleteness} 

 In contrast to the generic incompleteness accompanying the lowest bins of the
 luminosity function, we have another source of incompleteness attached to 
  the spectroscopic functionality of the grism images. The inclusion of a detected galaxy
   in the luminosity function depends on the detection of
       the  Ly$\alpha$ line.
     This source of incompleteness, which has no reasons to be confined to the
       lowest luminosity bins, can take two aspects.
 First, a number of features 
      with small equivalent width may be  missed by lack of contrast
      over the continuum flux: the distribution of EWs 
      (similar to the distribution of rest-frame EW in Fig. \ref{fig3})
      shows this happens below approximately EW = 20 \AA. Second, objects with
      EW larger than about 20 \AA\ may be  missed because their continuum flux
      is too low 
      for a spectrum to have been extracted:  Fig. \ref{fig2} shows that
      spectra are not systematically extracted below a FUV magnitude of 21.5.

       In order to understand the mechanisms of this second aspect of the
      incompleteness, we have illustrated
      the interplay between EW and observed FUV magnitudes
        in  Fig. \ref{fig8}. 
      Because of the limited redshift range, each luminosity bin 
      corresponds to a relatively narrow domain in this diagram.
      As an example,  Fig. \ref{fig8} shows the domain for
      the luminosity bin  log(L$_{Ly\alpha}$) 
      41.8 -- 42.0 (erg s$^{-1}$), between the two curves corresponding to
      Ly$\alpha$ line fluxes of  1.5 and 8.7  $\times 10^{-15}$ erg cm$^{-2}$
      s$^{-1}$. This domain is cut
      in two by the horizontal continuum flux limit  at magnitude 21.5.
      In order to quantify the incompleteness, we first calculate the 
      number of galaxies per 0.5 mag bin expected in our volume space as a
      function of UV magnitude.
      We use  the 
      luminosity function obtained with the FUV band of {\it GALEX} by 
       \citet{arn05} in the range $0.2 < z < 0.4$. In each 0.5 mag bin the
      galaxies are in turn distributed in   10 \AA\  EW bins according to  
      the observed EW distribution (above the limit of 20 \AA). 
      This results in a number of Ly$\alpha$ emitting galaxies
      for each elementary cell
      of size 0.5 mag and   10 \AA\  EW in  the diagram   of Fig. \ref{fig8}. 
      These galaxy numbers per elementary cell can be summed over
       the domain defined above
      in Fig. \ref{fig8}.
      The incompleteness factor is then taken 
       as the ratio of the sum over the
      entire  domain to the sum over the  domain above the limit of 21.5 in
      magnitude. If we account for the  galaxies fainter than 21.5 already contributing to  the space       densities   plotted in Fig. \ref{fig7}, we end up with an incompleteness factor of 6 
       for the Ly$\alpha$ luminosity bin   41.8 -- 42.0 (log erg s$^{-1}$) taken as example in  Fig. \ref{fig8}.

      We emphasize this correction factor does not result from
      simulated data but only from an evaluation using the 
     FUV luminosity function  and relying on the
      assumption that the observed EW distribution of galaxies
      apply to the less luminous galaxies involved in the evaluation.
      This assumption is supported by two facts. First, the
       observed EW distribution does not seem to change as a function of the UV luminosity 
       (Fig. \ref{fig5}) at least in the observed range of  luminosity. Second, the 
       bulk of the correction factor originates from galaxies 1 or 1.5 mag fainter than the 
       21.5 limit with EWs between 40 and 60  \AA. This range of EW values  seems less 
       prone to change with luminosity than extreme values.
         As the uncertainties on the correction factor remain severe and the amplitude
            of the correction is
      increasing for the low luminosity bins, we have repeated the evaluation only
      for the three luminosity bins brighter than (41.8 -- 42.),
      finding  correction factors
      of 4.2, 2.6, and 2.7 respectively. The brightest bin, corresponding to just one object, 
      has not been corrected for. 
      The space densities resulting from the  incompleteness correction are
      plotted in  Fig. \ref{fig7}.

         Fig. \ref{fig8} can also illustrate the other source of incompleteness resulting from 
         small EW features undetected by lack
         of contrast over the continuum; the relevant objects would lie
         in the domain defined by
          the two lines and the limit at EW $<$ 20 \AA. The EW distribution is
         essentially unknown in this area but, for all reasonable assumptions,
         an evaluation  based on the
         same scheme as above leads
         to a negligible additional factor.
          This is due to
         the fact that galaxies are bright in this domain (at least those
         related to the high luminosity bins) and consequently much  less dense 
         than the faint galaxies  involved in the first source of incompleteness described above.

\subsection{Comparisons with H$\alpha$ luminosity functions} 

  In the same way as the comparison of the Ly$\alpha$ luminosity  with the H$\alpha$ luminosity of an 
  individual star-forming galaxy places constraints on the  escape of Ly$\alpha$ photons through 
  the resonant scattering process, we may compare the Ly$\alpha$ and the
 H$\alpha$ luminosity functions of galaxies. Since the  escape of  Ly$\alpha$
  emission is expected to be highly variable  from galaxy to galaxy, the
  comparison will lead to an average (Ly$\alpha$ flux weighted) value of the escape, 
 as if all galaxies were the same. 
 For this comparison we have used the 
  H$\alpha$ luminosity function of  \citet{tre98} obtained at z $\sim 0.2$, close to our redshift 
  window. This  luminosity function is itself 
   consistent with other determinations  \citep{tre02,fuj03,nak04}. We adopt
  the values log($L^{*}$) = 41.92 ergs s$^{-1}$
  and log($\Phi^{*}$) = $-2.56$ Mpc$^{-3}$ as updated from   \citet{tre98} to
  current cosmology ($H_0 = 70$ km s$^{-1}$ Mpc$^{-1}$, $\Omega_{\Lambda} =
  0.7$,  $\Omega_{m} = 0.3$)  by \citet{fuj03}.

\begin{figure}
\plotone{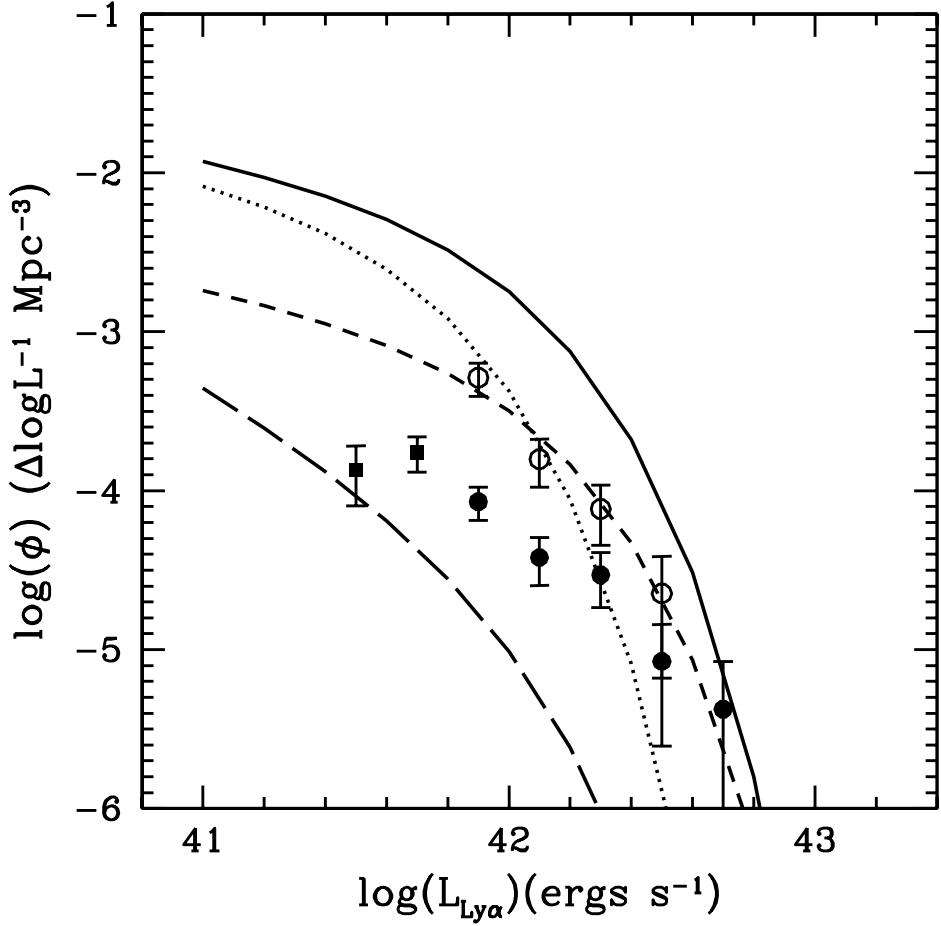}
\caption{The  space density of Ly$\alpha$ emitting galaxies  ($ 0.2 < z < 0.35$) per log  Ly$\alpha$ luminosity:
as measured (solid circles: all five fields, solid squares: CDFS, GROTH and NGPDWS fields), with an evaluation accounting for incompleteness
(open circles).   The error bars are Poisson errors in our bins. 
 Luminosity functions derived 
 from the H$\alpha$ luminosity function of   \citet{tre98} are shown with Ly$\alpha$/H$\alpha$ = 1
 for all galaxies (solid line);  Ly$\alpha$/H$\alpha$ = 0.5 for all galaxies
 (dotted line); and a least-square fit close to
  Ly$\alpha$/H$\alpha$ = 1 in 15  \% of galaxies (short-dashed). An evaluation of
  the Ly$\alpha$ luminosity function of narrow-line AGNs, derived from  the H$\alpha$ luminosity function of   \citet{hao05b}
  is displayed (long-dashed) for comparison.
\label{fig7}} 
\end{figure}

\begin{figure}
\plotone{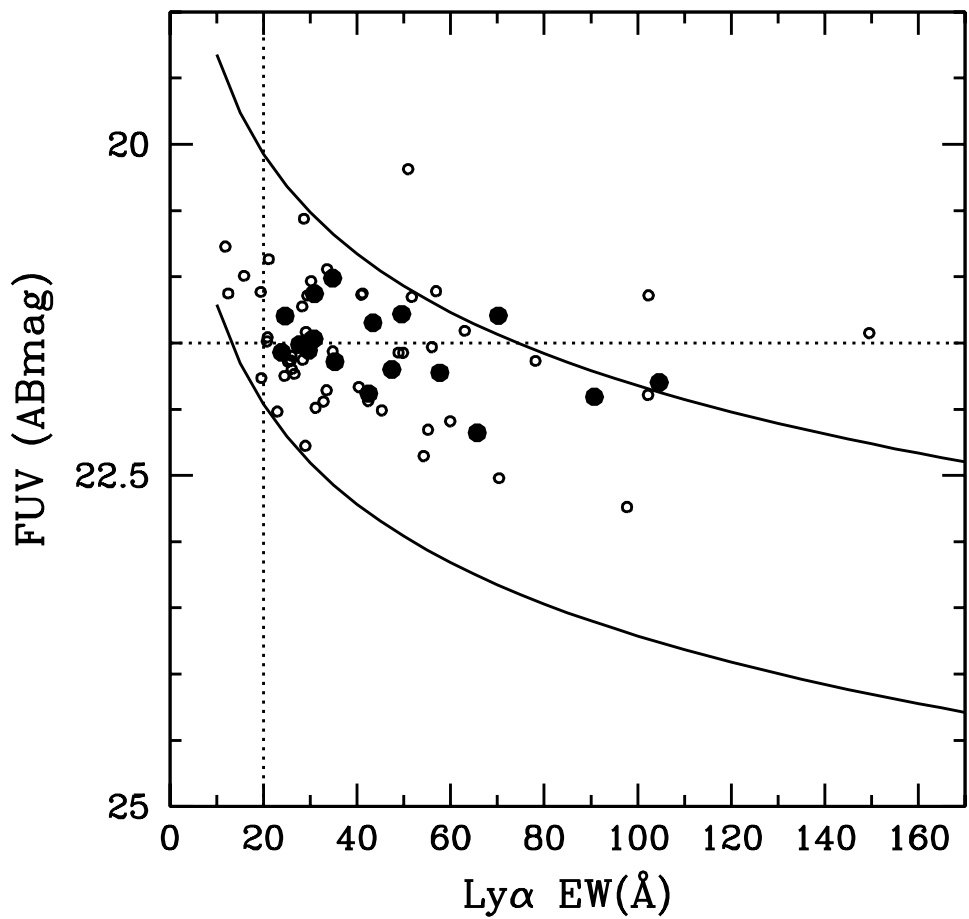}
\caption{ The FUV (AB mag) flux vs. the Ly$\alpha$ equivalent width. The
  objects from the  Ly$\alpha$ luminosity bin 41.8 - 42.0 (log(Ly$\alpha$)
  in ergs s$^{-1}$) lie between the two diagonal curves. In this domain, 
  objects below the horizontal line at mag 21.5 may be missed 
  because their continuum is too weak and their spectra  not extracted;
  objects at the left of the vertical line may be missed because 
  of a lack of contrast over the continuum. An evaluation of the 
  resulting incompleteness is described in section 4.2. 
  The observed data are overlaid (circles);  the
  filled circles are the 18 objects of the luminosity bin 41.8 -- 42.0.
  Two of these objects are slightly off the domain because of 
  the average calibration factor involved into the calculation of their 
   Ly$\alpha$ luminosity.
   \label{fig8}}
\end{figure}

  Because our binned data points are few and related to high luminosity values, we 
  assume  for our  Ly$\alpha$ luminosity function the same value 
  $\alpha  = - 1.35 $ as determined for the  H$\alpha$ LF. 
 The Ly$\alpha$/H$\alpha$ ratio and 
 the fraction of galaxies with  Ly$\alpha$
  emission are directly given by the factor of modifications of
 the  parameters  $L^{*}$ and  $\Phi^{*}$ to fit our Ly$\alpha$ data. A change
  of  $L^{*}$ alone does not work well (see one example in  Fig. \ref{fig7})
  and a  reasonable fit requires to decrease  $\Phi^{*}$,
  or, as currently observed, to have  Ly$\alpha$ in emission in only a
  fraction of the galaxies. Although systematic errors may be present in our
  completeness correction, we have carried out a least-square fit on the 5
  brightest luminosity data with  $L^{*}$ and  $\Phi^{*}$ as free
  parameters. The fit (Fig. \ref{fig7}), given by log($L^{*}$) = 41.98 $\pm0.09$ ergs s$^{-1}$
  and log($\Phi^{*}$) = $-3.40$ $\pm0.16$ Mpc$^{-3}$, implies 
  a Ly$\alpha$/H$\alpha$ ratio 
    of about 1 in  15 \% of the galaxies.
 As expected, the  Ly$\alpha$/H$\alpha$ ratio is
   much lower than the ratio of 8.7  predicted by the 
   case B recombination theory. This is in rough agreement with 
   the range of values reported in nearby galaxies by \citet{gia96}.
   This implies an average  Ly$\alpha$ escape fraction of about 0.1 somewhere between 
   the average values of 0.02 and 0.8 in the galaxy formation models of 
 \citet{led06}  and \citet{kob07} 
 respectively.

   At this stage, it is also possible to show  that the  Ly$\alpha$ luminosity function is not affected by the fact 
   that we have been unable to distinguish and remove narrow-line AGNs from our 
   sample of  Ly$\alpha$ emitting galaxy candidates.  We use the H$\alpha$  luminosity function
   of narrow-line AGNs derived from the SDSS by \citet{hao05b} with  the
criterion of  \citet{kau03} 
   which gives the largest number of objects.
   The Ly$\alpha$ LF is derived with the Ly$\alpha$/H$\alpha$ ratio of 3.24 
   reported by  \citet{vand01} from SDSS composite quasar spectra. The plot in Fig. \ref{fig7}
   shows that the contribution of narrow-line AGNs remains small even if we account for 
   some evolution between the redshift window of the SDSS sample (0 - 0.15) and ours (0.2 - 0.35).

\subsection{Comparisons with Ly$\alpha$ luminosity functions at high z} 
      
\begin{figure}
\plotone{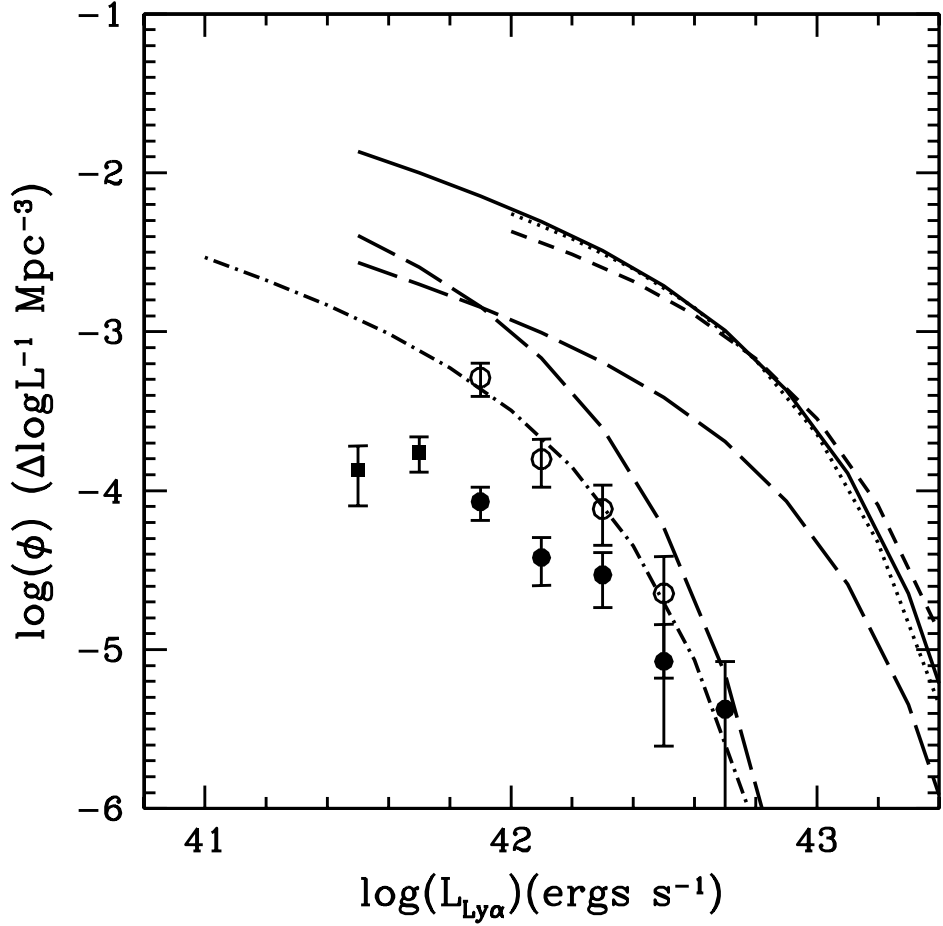}
\caption{The  space density of Ly$\alpha$ emitting galaxies ($ 0.2 < z < 0.35$) per log  Ly$\alpha$ luminosity:
as measured (solid circles: all five fields, solid squares: CDFS, GROTH and
NGPDWS fields),  
with an evaluation accounting for incompleteness
(open circles). Comparison  
 with  Ly$\alpha$ luminosity functions  at high redshifts: solid line,  \citet{vanb05} at $2.3 < z < 4.6$
 ( $L^{*}$ =  $5  \times 10^{42}$ ergs s$^{-1}$,  $\Phi^{*}$ =  0.0012 Mpc$^{-3}$, $\alpha = -1.6$); 
dotted line, \citet{gro07} at $z \sim 3.1$; 
 short-dashed line,  \citet{ouc07} at $z \sim 3.1$.
The dot-dashed LF is derived from a least-square fit on the 5 brightest points.
  The long-dashed lines show the impact of a factor 
 5 decrease of $L^{*}$ (the nearest curve to the data points) or  $\Phi^{*}$ in the LF of  \citet{vanb05},
 this factor corresponding
 to the decrease of  the UV luminosity density from   $z \sim 3$ to  $z \sim 0.3$. 
\label{fig9}}
\end{figure}

     The  space densities of  {\it GALEX} Ly$\alpha$ emitting galaxies 
  at $0.2 < z < 0.35$ are interesting to compare 
     with the  Ly$\alpha$ LFs found at high redshift. 
In  Fig. \ref{fig9}  we have  first  plotted
   the Ly$\alpha$ LF obtained by \citet{vanb05} at  redshifts $2.3 < z < 4.6$
  which is selected because it  is the closest from our data in terms of
     redshift. 
 In addition, the authors
   have  compared  their data  with  the measurements available
 at the time (references therein and their Figure 5 ). 
They conclude that the luminosity function of  Ly$\alpha$ emitters does not 
   significantly change from z $\sim 3.4$ to z = 5.7. This  is  confirmed by more recent 
   determinations  from   \citet{aji06}, 
 \citet{tap06},  \citet{shi06},  \citet{mur07},  \citet{gro07},
    \citet{ouc07} and  \citet{daw07}. 
A   few LFs at $z >  6$ have been left out of the comparison because 
 variations of  Ly$\alpha$ LFs are possible that would not be related to the galaxies 
  themselves but  to the IGM opacity resulting from
   changes in the IGM neutral fraction accompanying the re-ionisation.
   Of the recent  Ly$\alpha$ LFs at high redshift,
we have elected for clarity to reproduce  in  Fig. \ref{fig9} only those closest from our redshift range,
i.e. the two from   \citet{gro07} and   \citet{ouc07} at $ z = 3.1 $. 
 
 Fig. \ref{fig9} shows that  the space density of the  Ly$\alpha$ emitting
 galaxies 
is much lower in the range  $0.2 < z < 0.35$ than at redshifts  about 3.
We have quantified this factor by
a least-square fit on the 5 brightest data points, using the same  $\alpha$
 parameter ($ -1.6$) as  \citet{vanb05} 
  (adopted by these authors from  \citet{ste99}); we have obtained
 log($L^{*}$) = 42.03 $\pm0.08$ ergs s$^{-1}$
  and log($\Phi^{*}$) = $-3.47$ $\pm0.17$ Mpc$^{-3}$. 

 By comparison with the values $L^{*}$ =  $5 \pm 1.8   \times 10^{42}$ ergs
   s$^{-1}$ 
and  $\Phi^{*}$ = $ 0.0012 \pm 0.0005$   Mpc$^{-3}$  of  \citet{vanb05}, our determination implies  
 a  Ly$\alpha$ luminosity density (LD) $\sim 16.5$  times larger at  $z \sim 3$  than at
$0.2 < z < 0.35$.
In comparison, the FUV LD of  \citet{sch05} increases by a factor of about 5 between 
$z = 0.3$ and  $z \sim 3$, an  increase  consistent with the compilations of SFR evolution 
of  \citet{hop04} and  \citet{hop06}. We have illustrated 
this difference in the evolution rates of the 
FUV and Ly$\alpha$ luminosity densities in Fig. \ref{fig9} 
  by plotting the effect of a factor 5 decrease of the parameters  $L^{*}$ or
   $\Phi^{*}$  of the LF of  \citet{vanb05}. Although the effect looks significant,
 we discuss further 
 the  uncertainties in our approach as well as possible interpretations.

1) Insofar as the FUV LD measures the evolution of the massive stellar content 
of the galaxies, the faster rate of evolution of the  Ly$\alpha$ LD suggests a real increase 
(by a factor of about 3) of the 
 Ly$\alpha$ escape fraction from $z  \sim  0.3$ to  $z \sim  3$. This is consistent with 
 the observations of an increase with redshift of the fraction of  (EW $>$ 20) Ly$\alpha$ emitting galaxies 
 (this paper,  \citet{red07}). This is also consistent with the current idea that galactic winds are increasing
 with redshift,  favouring the  Ly$\alpha$ escape while mitigating the
 increased fraction of neutral hydrogen in galaxies at high redshift.

2) The rate of increase of the  Ly$\alpha$ LD between 
$z = 0.3$ and  $z \sim 3$ depends on the uncertainties of 
 $L^{*}$ and
   $\Phi^{*}$ at both redshifts, i.e. 4 parameters. 
If we combine the variances obtained from our
least-square fit with those given by  \citet{vanb05}, the standard deviation
on the factor 16.5 is as large as 11.5 and the  evolution of the  
 Ly$\alpha$ LD  is faster than the FUV LD at a significance  of about 84\%.   

 3)  A change in the faint-end slope $\alpha$ of the LF  may play a role in the 
  evolution of the  Ly$\alpha$ LD which writes as $L^{*} \Phi^{*}  \Gamma(\alpha + 2)$.
  The parameter $\alpha$ is indeed poorly constrained by the space densities
  of the luminous  Ly$\alpha$ emitters that are observed both by {\it GALEX} and at
  high redshift. This parameter
   is  assumed to be the same in our comparison between low and high
  redshifts. 
A steepening of the slope
 $\alpha$  at low z, resulting in an increase of  $\Gamma(\alpha + 2)$ could make the 
  evolution of the  Ly$\alpha$ LD matching that of the FUV LD. This would mean an increase 
   of the  Ly$\alpha$ escape fraction in low-luminosity objects at low z. 
  Such a variation, however, would be opposite   to the current trend 
  of a steepening of the faint-end slope with redshift (e.g.  \citet{rya07} for FUV LF) as expected 
  in the hierarchical formation scenario of galaxies.

    In conclusion, the faster rate of evolution of the  Ly$\alpha$ luminosity
  density with respect to the cosmic star formation rate is significant at  
   about 84\% and
 indicates probably an
  increase of the average Ly$\alpha$ escape fraction from  $z  \sim  0.3$ to  $z \sim  3$.

\section{Conclusion}

     A blind search of potential  Ly$\alpha$ emission features has been
     conducted on 7018 spectra obtained in 5 deep spectroscopic exposures with
     {\it GALEX}.
      96  Ly$\alpha$ emitting galaxy candidates are identified, after a
     separation from AGNs essentially on the basis of the FWHM. They are 
      mostly in the reshift range (0.2 $-$ 0.35). 
     The following properties stand out:

     1) The Ly$\alpha$ rest frame equivalent width distribution is
        comparable with that reported by  \citet{sha03}  for LBGs at   $z \sim
        3$. The   fraction of galaxies with
      Ly$\alpha$ emission (EW $>  20$ \AA) seems smaller, 15\%
        against 25\% in the LBG sample. 

     2) No trend is found between the EW and the (FUV $-$ NUV) color, in 
        agreement with a decoupling of the reddening of line and continuum
        photons in  Ly$\alpha$ resonant scattering. There is no trend either of the
        EW  with the UV luminosity, except a decrease in a sub-sample of 
   ultra-violet luminous galaxies (UVLGs). As at very high-redshifts,
        small-number statistics
                might play a role here. 
    A larger fraction of  Ly$\alpha$ emitting galaxies (EW $>
        20$ \AA) 
            is found among UVLGs. 

     3) A sub-sample of 66 emission features of better quality
       and  strictly in the  reshift range (0.2 $-$ 0.35)
       has been used to calculate the space densities of the  Ly$\alpha$ emitting
       galaxies. 
  A scheme has been presented
  to correct for a major source of incompleteness, the fact that spectra of objects with
  significant EW may have not been extracted because their continuum is too weak. 
     
    4) A comparison with the  H$\alpha$  luminosity function of  \citet{tre98}
       in the same redshift domain is consistent with an average  Ly$\alpha$
       to  H$\alpha$ ratio of 1 in about  15\% of the galaxies.

    5) A comparison of the  Ly$\alpha$  luminosity functions at 
  $z  \sim 0.3$ and   $z \sim 3$ shows 
     an evolution beyond that expected from the evolution of 
    the massive stellar content of star-forming 
     galaxies at a significance level of  84\%, suggesting a  increase  of the average  Ly$\alpha$ escape
  fraction  with redshift.



\acknowledgments

{\it GALEX (Galaxy Evolution Explorer)} is a NASA Small Explorer, launched in April 2003.
We gratefully acknowledge NASA's support for construction, operation,
and science analysis for the {\it GALEX} mission,
developed in cooperation with the Centre National d'Etudes Spatiales
of France and the Korean Ministry of 
Science and Technology. 
This research has made use of the NASA/IPAC Extragalactic Database (NED) 
which is operated by the Jet Propulsion Laboratory, California Institute of
Technology,
under contract with the National Aeronautics and Space Administration.

\clearpage




\clearpage

\clearpage

\begin{table}
\begin{center}
\caption{Characteristics of the five {\it GALEX} spectroscopic fields used.\label{tbl-1}}

\begin{tabular}{lrrrrr}

\tableline\tableline 
           & CDFS-00  & ELAISS1-00 & GROTH-00 & NGPDWS-00 & SIRTFFL-00  \\

\tableline
Exposure time (s) & 149315 & 84086 & 281713 & 139598 & 79616\\
Center of field RA\tablenotemark{a} & 53.128 & 9.638 & 214.992 & 219.156 & 259.124\\
Center of field DEC\tablenotemark{a} & -27.871 &  -43.990 & 52.782 & 35.171 & 59.909 \\
Identified sources\tablenotemark{b} & 38697 & 30129  & 43545 & 40619 & 28306   \\
Extracted spectra &  1419         &  925    & 2028  & 1202         & 1444 \\
Emission features &   22  & 9 & 39 &       19   &         7\\
Used for the LF\tablenotemark{c} & 15  & 5&  29 & 11  & 6 \\
\tableline

\end{tabular}

\tablenotetext{a}{in decimal degrees.} 
\tablenotetext{b}{entries in the matched catalog of the FUV and NUV images.}
\tablenotetext{c}{used for the evaluation of the luminosity function }

\end{center}
\end{table}

\clearpage

\begin {deluxetable}{llcccccccccc}
\tabletypesize{\scriptsize}
\tablecaption{The Ly$\alpha$ emitting galaxy candidates \label{tbl-2}}
\tablewidth{0pt}
\tablehead{
\colhead{Field} & 
\colhead{ID} & 
\colhead{RA} & 
\colhead{DEC} & 
\colhead{EW} &
\colhead{Q} & 
\colhead{Ly$\alpha$ Flux} & 
\colhead{FUV} & 
\colhead{COLOR} &
\colhead{z} &
\colhead{L(Ly$\alpha$)}&
\colhead{z0}\\
\colhead{(1)} & 
\colhead{(2)} &   
\colhead{(3)} & 
\colhead{(4)} &  
\colhead{(5)} & 
\colhead{(6)} &  
\colhead{(7)} & 
\colhead{(8)} &
\colhead{(9)} & 
\colhead{(10)} &
\colhead{(11)} &
\colhead{(12)}
}

\startdata
CDFS &  1348 & 53.2405 & -28.3883 & 43.4 & 1 & 5.43 & 21.35 & 0.615 & 0.217 & 41.87&  \\
CDFS &  1821 & 53.2585 & -28.3577 & 24.6 & 1 & 3.37 & 21.30 & 0.303 & 0.251 & 41.81&  \\
CDFS &  2422 & 52.8947 & -28.3395 & 14.4 & 2 & 4.79 & 20.54 & 0.488 & 0.176 & 41.63& 0.1728(1) \\
CDFS &  3801 & 52.7375 & -28.2794 & 29.7 & 1 & 3.62 & 21.55 & 0.528 & 0.285 & 41.97&    \\
CDFS &  4927 & 52.9765 & -28.2386 & 102. & 1 & 8.12 & 21.89 & 0.450 & 0.283 & 42.31&   \\
CDFS &  5007 & 53.5412 & -28.2554 & 97.7 & 2 & 3.14 & 22.74 & 1.283 & 0.344 & 42.09&  \\
CDFS &  5448 & 53.0780 & -28.2224 & 149. & 1 & 11.4 & 21.42 & 0.312 & 0.283 & 42.46&  \\
CDFS &  6523 & 53.0616 & -28.1865 & 47.4 & 1 & 3.48 & 21.70 & 0.551 & 0.264 & 41.87&  \\
CDFS &  6535 & 52.9622 & -28.1890 & 34.8 & 1 & 5.16 & 21.01 & 0.386 & 0.216 & 41.85&  \\
CDFS &  6617 & 53.1743 & -28.1903 & 50.9 & 1 & 17.2 & 20.19 &-0.037 & 0.208 & 42.33&   \\
CDFS &  7100 & 52.9993 & -28.1644 & 29.1 & 1 & 2.96 & 21.42 & 0.282 & 0.239 & 41.71&  \\
CDFS & 10526 & 53.5868 & -28.0657 & 24.3 & 1 & 1.75 & 22.11 & 0.940 & 0.361 & 41.89&  \\ 
CDFS & 10937 & 53.7850 & -28.0454 & 63.0 & 1 & 5.89 & 21.41 & 0.423 & 0.346 & 42.37&  \\
CDFS & 11518 & 53.0498 & -28.0250 & 57.7 & 1 & 5.50 & 21.72 & 0.249 & 0.218 & 41.89& 0.212(2) \\
CDFS & 16104 & 53.2360 & -27.8879 & 23.2 & 2 & 4.35 & 21.14 & 0.544 & 0.374 & 42.32& 0.365(3) \\
CDFS & 17033 & 52.7601 & -27.8584 & 38.1 & 3 & 3.57 & 21.96 & 0.494 & 0.340 & 42.14&  \\
CDFS & 18142 & 52.8861 & -27.8344 & 24.4 & 2 & 4.63 & 21.04 & 0.052 & 0.183 & 41.64& 0.133(2) \\
CDFS & 19355 & 53.7296 & -27.8008 & 28.7 & 1 & 7.49 & 20.56 & 0.783 & 0.314 & 42.38&  \\
CDFS & 21667 & 53.2803 & -27.7424 & 11.8 & 2 & 2.50 & 20.77 & 0.506 & 0.219 & 41.55& 0.216(4)  \\
CDFS & 21739 & 53.7113 & -27.7293 & 34.8 & 1 & 4.05 & 21.56 & 0.731 & 0.323 & 42.14&  \\
CDFS & 30899 & 53.3592 & -27.4543 & 78.6 & 1 & 7.29 & 21.63 & 0.502 & 0.352 & 42.48& \\
CDFS & 33311 & 53.1045 & -27.2904 & 58.0 & 1 & 8.65 & 21.43 & 0.271 & 0.391 & 42.66&  \\ 

ELAISS1 & 13715 &  9.6383 & -44.0090 & 21.9 & 3 &  2.93 & 21.64 &  1.308 & 0.213 & 41.59 &\\ 
ELAISS1 & 16998 &  9.5205 & -43.8745 & 56.9 & 1 &  8.16 & 21.11 &  0.508 & 0.223 & 42.08 &\\
ELAISS1 &  6587 &  9.5590 & -44.2436 & 102. & 1 &  17.9 & 21.14 & -0.107 & 0.272 & 42.62 &\\
ELAISS1 &  8180 &  9.8839 & -44.1917 & 38.2 & 2 &  10.1 & 20.34 &  0.881 & 0.188 & 42.01 & 0.1862(5)\\
ELAISS1 & 21062 &  9.6663 & -43.7225 & 12.5 & 2 &  2.81 & 21.13 &  0.204 & 0.211 & 41.56 &\\
ELAISS1 & 23257 &  9.4752 & -43.6410 & 32.7 & 3 &  5.17 & 21.34 &  0.512 & 0.294 & 42.15 &\\
ELAISS1 & 23425 &  9.3711 & -43.6356 & 21.1 & 2 &  6.49 & 20.87 &  0.387 & 0.300 & 42.27 &\\
ELAISS1 & 16921 & 10.2733 & -43.8748 & 20.3 & 3 &  4.14 & 21.53 &  0.929 & 0.312 & 42.12 &\\
ELAISS1 &  2386 & 10.0078 & -44.4288 & 32.8 & 2 &  2.81 & 21.94 &  0.732 & 0.268 & 41.80 &\\
GROTH &  6834 & 215.6564 & 52.4520 &  11.1 & 2 &  2.01 & 21.26 & 0.466 & 0.197 & 41.35 &\\
GROTH & 32462 & 215.1704 & 53.1138 &  20.8 & 2 &  1.61 & 21.49 & 0.442 & 0.202 & 41.28 & 0.2004(6)\\
GROTH & 36896 & 214.9730 & 53.3764 &  29.4 & 1 &  3.29 & 21.14 & 0.373 & 0.199 & 41.57 &\\
GROTH &  7430 & 214.4311 & 52.4683 & 104.6 & 1 &  7.67 & 21.80 & -0.101& 0.211 & 42.00 & 0.2092(7)\\
GROTH &  5087 & 214.5594 & 52.3956 &  44.6 & 3 &  1.23 & 22.61 & 0.746 & 0.215 & 41.22 &\\
GROTH & 34512 & 214.2955 & 53.1980 &  70.2 & 1 &  5.44 & 21.29 & 2.580 & 0.215 & 41.87 & 0.2139(6)\\
GROTH &  8885 & 215.6107 & 52.5075 &  42.4 & 2 &  2.45 & 21.94 & 0.379 & 0.221 & 41.55 &\\
GROTH &  2368 & 214.5933 & 52.3067 &  54.2 & 2 &  1.89 & 22.35 & 0.655 & 0.242 & 41.52 &\\
GROTH & 18322 & 214.5216 & 52.7522 &  25.8 & 2 &  2.78 & 21.63 & 0.335 & 0.247 & 41.71 & 0.24438(8) \\
GROTH &  2682 & 214.7013 & 52.2986 &  19.3 & 3 &  1.52 & 22.03 & 0.370 & 0.244 & 41.44 &\\
GROTH &  5715 & 214.2262 & 52.4111 &  27.7 & 1 &  3.59 & 21.51 & 0.507 & 0.250 & 41.83 & 0.24678(8)\\
GROTH & 19002 & 214.4387 & 52.7719 &  59.9 & 1 &  2.94 & 22.09 & 0.265 & 0.248 & 41.74 & 0.24419(8)\\
GROTH & 17005 & 215.1805 & 52.7188 &  26.0 & 2 &  2.87 & 21.59 & 0.304 & 0.252 & 41.74 &\\
GROTH &  4719 & 214.8114 & 52.3908 &  25.3 & 3 &  1.42 & 22.16 & 0.394 & 0.257 & 41.46 &\\
GROTH & 20285 & 215.1330 & 52.7994 &  30.9 & 1 &  4.11 & 21.13 & 0.401 & 0.256 & 41.92 &\\
GROTH & 21404 & 215.1861 & 52.8351 &  25.3 & 1 &  2.53 & 21.64 & 0.290 & 0.256 & 41.70 &\\
GROTH & 12279 & 214.3008 & 52.5991 &  19.6 & 2 &  2.08 & 21.76 & 0.527 & 0.264 & 41.65 & 0.26113(8)\\
GROTH & 14069 & 215.3526 & 52.6555 &  29.0 & 2 &  1.80 & 22.28 & 0.463 & 0.260 & 41.57 &\\
GROTH & 21024 & 214.7318 & 52.8245 &  41.0 & 2 &  2.01 & 21.13 &-0.026 & 0.269 & 41.65 & 0.2633(6)\\
GROTH & 36336 & 214.5818 & 53.3393 &  40.3 & 3 &  2.24 & 22.29 & 1.265 & 0.268 & 41.70 &\\
GROTH & 37457 & 214.7951 & 53.2660 &  90.7 & 1 &  4.38 & 21.90 & 0.925 & 0.266 & 41.98 &\\
GROTH &  3488 & 214.9704 & 52.3502 &  40.4 & 2 &  2.52 & 21.83 & 0.541 & 0.269 & 41.75 &\\
GROTH & 37380 & 215.1904 & 53.3248 &  26.1 & 2 &  1.93 & 21.70 & 0.472 & 0.269 & 41.64  &\\
GROTH &  3525 & 214.7796 & 52.3522 &  55.2 & 2 &  2.41 & 22.15 & 0.766 & 0.271 & 41.74  &\\
GROTH & 29573 & 214.8762 & 53.0349 &  23.9 & 3 &  1.32 & 22.05 & 0.586 & 0.271 & 41.48 &\\
GROTH & 31403 & 214.2910 & 53.0867 &  33.5 & 1 &  2.25 & 21.86 & 0.570 & 0.270 & 41.71  &\\
GROTH & 33559 & 214.9025 & 53.1601 &  23.0 & 2 &  1.77 & 22.02 & 0.377 & 0.273 & 41.61 &\\
GROTH & 17867 & 215.8429 & 52.7425 &  22.0 & 3 &  1.21 & 22.36 & 0.621 & 0.282 & 41.48 &\\
GROTH & 17525 & 215.8241 & 52.7135 &  31.9 & 3 &  2.62 & 21.77 & 0.756 & 0.283 & 41.82 &\\
GROTH &  9045 & 214.9070 & 52.5070 &  24.5 & 2 &  1.86 & 21.75 & 0.475 & 0.286 & 41.68 &\\
GROTH & 15686 & 215.9047 & 52.6719 &  26.6 & 2 &  2.21 & 21.73 & 0.461 & 0.287 & 41.76 &\\
GROTH & 13305 & 215.8867 & 52.6237 &  24.6 & 3 &  1.70 & 21.96 & 0.471 & 0.286 & 41.64 &\\
GROTH & 21579 & 214.2081 & 52.8388 &  31.1 & 2 &  1.71 & 21.99 & 0.501 & 0.287 & 41.65 &\\
GROTH & 28751 & 214.7328 & 52.9926 &  42.5 & 1 &  2.92 & 21.88 & 0.509 & 0.290 & 41.89 &\\
GROTH & 23096 & 215.5339 & 52.8738 &  70.4 & 1 &  3.80 & 22.52 & 1.220 & 0.307 & 42.06 &\\
GROTH & 29558 & 214.0695 & 53.0259 &  65.7 & 2 &  2.58 & 22.18 & 1.077 & 0.329 & 41.96 &\\
GROTH & 19364 & 215.7761 & 52.7797 &  45.3 & 1 &  2.67 & 22.01 & 0.681 & 0.347 & 42.04 &\\
GROTH &  5549 & 215.4769 & 52.4065 &  28.3 & 2 &  4.24 & 21.22 & 0.832 & 0.350 & 42.24 &\\
GROTH & 10182 & 214.3223 & 52.5384 &  55.7 & 3 &  4.89 & 21.85 & 0.354 & 0.461 & 42.59 &\\

NGPDWS & 28760 & 219.1979 & 35.4351 & 115.0 & 1 & 20.66 & 20.93 & 0.061 & 0.180 & 42.27 &\\
NGPDWS & 23216 & 218.6954 & 35.2844 &  34.0 & 1 &  4.09 & 21.42 & 0.543 & 0.192 & 41.63 &\\
NGPDWS & 32840 & 219.2433 & 35.5977 &  20.9 & 2 &  2.32 & 21.46 & 0.771 & 0.209 & 41.47 &\\
NGPDWS & 11927 & 219.1004 & 34.9935 &  48.9 & 1 &  4.34 & 21.57 & 0.267 & 0.216 & 41.77  &\\
NGPDWS & 19918 & 219.1990 & 35.1757 &  26.2 & 3 &  3.74 & 21.13 & 0.711 & 0.253 & 41.86 &\\
NGPDWS & 23690 & 219.8446 & 35.3075 &  41.2 & 1 &  5.88 & 21.12 & 0.150 & 0.250 & 42.05 &\\
NGPDWS & 28521 & 219.0262 & 35.4586 &  23.9 & 1 &  4.05 & 21.57 & 0.189 & 0.253 & 41.90 &\\
NGPDWS & 35813 & 219.0558 & 35.7291 &  30.4 & 2 &  3.70 & 21.47 & 0.084 & 0.263 & 41.90 &\\
NGPDWS & 33782 & 219.5770 & 35.6305 &  78.2 & 1 &  6.05 & 21.63 & 0.521 & 0.264 & 42.11 &\\
NGPDWS & 10002 & 219.0922 & 34.9421 &  30.8 & 1 &  4.33 & 21.47 & 0.312 & 0.272 & 42.00 &\\
NGPDWS &  6731 & 219.1529 & 34.8428 &  51.7 & 1 & 10.10 & 21.15 & 0.496 & 0.283 & 42.41 &\\
NGPDWS &  2111 & 219.3562 & 34.6855 &  35.3 & 1 &  3.67 & 21.64 & 0.641 & 0.292 & 42.00 &\\
NGPDWS & 30997 & 218.7306 & 35.5246 &  19.4 & 2 &  3.13 & 21.11 & 0.523 & 0.320 & 42.02 &\\
NGPDWS &  1133 & 219.1333 & 34.6415 & 153.0 & 3 &  4.86 & 22.46 & 1.074 & 0.358 & 42.33 &\\
NGPDWS & 27558 & 219.1931 & 35.4176 & 117.0 & 3 &  6.02 & 22.20 & 0.993 & 0.372 & 42.46 &\\
NGPDWS & 10713 & 219.7463 & 34.9603 &  69.3 & 3 &  5.16 & 21.94 & 0.647 & 0.374 & 42.40 &\\
NGPDWS &  6321 & 219.7806 & 34.8359 &  77.2 & 3 &  7.43 & 22.32 & 0.944 & 0.374 & 42.55 &\\
NGPDWS & 35880 & 219.2820 & 35.6919 & 161.0 & 3 &  4.69 & 22.79 & 1.859 & 0.468 & 42.58 &\\
NGPDWS &  4226 & 219.3340 & 34.7577 &  56.0 & 1 &  6.49 & 21.53 & 0.350 & 0.328 & 42.36 &\\   
SIRTFFL & 14450 & 259.2110 & 59.9642 & 41.8 & 1 & 22.19 & 19.77 & 0.133 & 0.185 & 42.33 &\\
SIRTFFL & 14297 & 258.6840 & 59.9474 & 28.4 & 2 &  2.52 & 21.62 & 0.494 & 0.219 & 41.55 &\\
SIRTFFL & 14085 & 258.1492 & 59.9468 & 15.8 & 2 &  2.87 & 20.99 & 0.624 & 0.225 & 41.63 &\\
SIRTFFL & 10895 & 258.5918 & 59.8333 & 49.5 & 1 &  6.18 & 21.28 & 0.192 & 0.233 & 42.00 &\\
SIRTFFL &   958 & 258.8205 & 59.3897 & 30.2 & 2 &  3.45 & 21.03 & 0.547 & 0.235 & 41.76 &\\
SIRTFFL &  4246 & 259.8606 & 59.5599 & 49.9 & 2 &  3.58 & 21.57 & 0.556 & 0.237 & 41.78 &\\
SIRTFFL &  2856 & 259.3800 & 59.4875 & 33.6 & 2 &  4.72 & 20.94 & 0.610 & 0.302 & 42.14 &\\

\enddata
\tablecomments{
Col.(5):  Ly$\alpha$  equivalent width (\AA).
Col.(6):  Quality of the detection (1 good; 2 fair; 3 uncertain).
Col.(7): Emission line flux in $10^{-15}$ erg cm$^{-2}$  s$^{-1}$. 
Col.(8): FUV magnitude (in the AB system) from {\it GALEX} photometry.
Col.(9): FUV-NUV color from {\it GALEX} photometry .
Col.(10): Redshift assuming that the emission feature is  Ly$\alpha$. 
Col.(11):  Ly$\alpha$  luminosity in  log (erg  s$^{-1}$). 
Col.(12): Redshift found in the literature (NED) and references: 
(1) 2dF (http://www.mso.anu.edu.au/2dFGRS/),
 (2) COMBO-17 \citet{wol04}, (3) \citet{van06},
 (4) SARS \citet{way05},   (5) LCRS  \citet{she96}, (6) SDSS, (7) DEEP1
Groth Strip \citet{wei06}, (8) DEEP2 Data Release 3 \citet{dav07}
}
\end{deluxetable}

 \clearpage



\clearpage
\end{document}